\documentclass[letterpaper]{article}
\usepackage{isea}
\usepackage[pdftex]{graphicx}
\usepackage{times}
\usepackage{helvet}
\usepackage{courier}
\usepackage{listings}
\usepackage{appendix}
\usepackage{hyperref}
\usepackage[numbers]{natbib}

\title{Evaluating BBRv2 on the Dropbox Edge Network}
\author{Alexey Ivanov\\
Dropbox, Inc\\
Mountain View, CA\\
SaveTheRbtz@GMail.com\\
\newline
\newline
}

\begin{document} 
\maketitle
\begin{abstract}
Nowadays, loss-based TCP congestion controls in general and CUBIC specifically became the \textit{de facto} standard for the Internet.  BBR congestion control challenges the loss-based approach by modeling the network based on estimated bandwidth and round-trip time.  At Dropbox, we've been using BBRv1 since 2017 and are accustomed to its pros and cons.  BBRv2 introduces a set of improvements to network modeling (explicit loss targets and inflight limits) and fairness (differential probing and headroom for new flows.) In this paper, we go over experimental data gathered on the Dropbox Edge Network.  We compare BBRv2 to BBRv1 and CUBIC showing that BBRv2 is a definite improvement over both of them.  We also show that BBRv2 experimental results match its theoretical design principles. 
\end{abstract}

\section{Keywords}
BBR, BBRv2, CUBIC, TCP, Congestion Control, AQM.

\section{Introduction}
Since the “Bottleneck Bandwidth and Round-trip propagation time” (BBR) congestion control paper\cite{Cardwell2017} was released it became production-ready and was added to Linux, FreeBSD, and Chrome (as a part of QUIC.)  Back then, Dropbox evaluated\cite{Ivanov2017} BBRv1 congestion control on our edge network and it showed promising results compared to CUBIC:

\begin{figure}[h]
\includegraphics[width=3.31in]{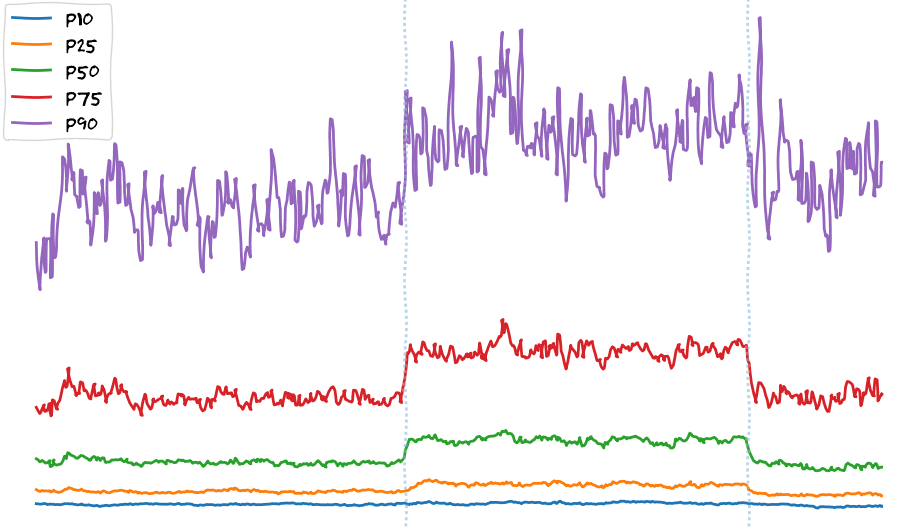}
\caption{Dropbox desktop clients’ download goodput.}
\label{fig:bbrv1goodput}
\end{figure}

After BBRv1 has been fully deployed to Dropbox Edge Network we started to identify its corner cases and shortcomings.  Some of them were eventually fixed, for example, BBRv1 being measurably slower for Wi-Fi users.  Other trade-offs were quite conceptual, for example BBRv1’s unfairness towards loss-based congestion controls\cite{2019arXiv190303852T}, RTT-unfairness between BBRv1 flows\cite{Hock2017}, and its disregard for packet loss.  Dropbox Edge Network \textbf{observed packet loss rates up to 6\% on hosts using BBRv1, compared to around 0.5\% on hosts using CUBIC with pacing}.

\section{BBRv1 shortcomings}

BBR developers were aware of these problems and actively worked on solutions. Following issues were identified\cite{ietf104}: 
\begin{itemize}
\item Low throughput for Reno/CUBIC flows sharing a bottleneck with bulk BBR flows.
\item Loss-agnostic; high loss if $bottleneck<1.5*BDP$.
\item ECN-agnostic.
\item Low throughput for paths with high degrees of aggregation (e.g. Wi-Fi.)
\item Throughput variation due to low cwnd in \texttt{PROBE\_RTT}.
\end{itemize}

\section{BBR version 2}

BBRv2 aims to solve some of the major drawbacks of the first version.  Here is the list of BBR design principles (\textbf{bold} means it’s new in BBRv2\cite{ietf105}.  See Table~\ref{tab:bbr2whatsnew}.):
\begin{itemize}
\item \textbf{Leave headroom: leave space for entering flows to grab.}
\item \textbf{React quickly: using loss/ECN, adapt to delivery process now to maintain flow balance.}
\item Don’t overreact: don’t do a multiplicative decrease on every round trip with loss/ECN.
\item \textbf{Probe deferentially: probe on a time scale to allow coexistence with Reno/CUBIC.}
\item Probe robustly: try to probe beyond estimated max bw, max volume before we cut estimation.
\item \textbf{Avoid overshooting: start probing at an inflight measured to be tolerable.}
\item Grow scalably: \textbf{start probing at 1 extra packet}; grow exponentially to use free capacity.
\end{itemize}

\begin{table*}
	\centering
	\setlength\tabcolsep{8pt}
	\renewcommand{\arraystretch}{1.2}
	\begin{tabular}{|p{3.3cm}|p{3.8cm}|p{3.3cm}|p{5.1cm}|}
        \hline
        & \textbf{Cubic} & \textbf{BBRv1} & \textbf{BBRv2} \\
        \hline\hline
        
		\textbf{Model parameters for the state machine} & N/A & Throughput, RTT & Throughput, RTT, max aggregation, max inflight \\ \hline
		
		\textbf{Loss}  & Reduce cwnd by $30\%$ on window by any loss &  N/A  &   Explicit loss rate target \\ \hline
		
		\textbf{ECN}   & RFC3168 (Classic ECN) & N/A   &   DCTCP-inspired ECN (See ``\nameref{sec:ecn}'' section.) \\ \hline
		
		\textbf{Startup} &  Slow-start until RTT rises (Hystart) or any loss  &  Slow-start until throughput plateaus & Slow-start until throughput plateaus or ECN or Loss rate > target \\ \hline
	\end{tabular}
  \caption{What’s new in BBRv2: a summary}
  \label{tab:bbr2whatsnew}
\end{table*}

\section{Test constraints}

\subsection{Non-mobile ISPs}
The test was focusing on the Dropbox Desktop Client traffic and therefore mostly excludes mobile ISPs.

\subsection{Bulk traffic focus}

This experiment was aimed at high-throughput workloads.  All TCP connections and nginx logs mentioned in the paper were filtered by having at least 1Mb of data transferred.  

\subsection{Localized test}

This experiment was performed in a single point of presence (PoP) in Tokyo, Japan.  Therefore, it carries over some of the biases common to traffic from that geographic area, including internet speeds, operating systems, client devices, etc.

\subsection{Real user traffic}

This is a real-world experiment with all of its pros and cons, including ISPs with heavy over-subscription, broken embedded TCP/IP implementations, etc\footnote{For testing congestion controls in a lab environment \texttt{github.com/google/transperf} can be used.  It allows testing TCP performance over a variety of emulated network scenarios, including RTT, bottleneck bandwidth, and policed rate that can change over time.}.

\section{Test setup}
Following combinations of kernels and congestion control algorithms were tested:
\begin{itemize}
\item 5.3 kernel, \texttt{cubic}
\item 5.3 kernel, \texttt{bbr}
\item 5.3 kernel, \texttt{bbr2}
\end{itemize} 

All of the servers are using a combination of \texttt{mq} and \texttt{sch\_fq} qdiscs with default settings.
Kernel is based on \texttt{Ubuntu-hwe-edge-5.3.0-18.19\_18.04.2} with all patches from the \texttt{v2alpha-2019-11-17} tag.

Graphs were generated from either connection-level information from \texttt{ss -neit} sampling, machine-level stats from \texttt{/proc}, or server-side application-level metrics from web-server logs.

\section{Caveats}

\subsection{Keeping the kernel up-to-date}
Newer kernels usually bring quite a bit of improvement\footnote{Recent Linux kernels also include mitigations for the newly discovered CPU vulnerabilities.  We highly discourage disabling them (especially on the edge!) so be prepared to also take a CPU usage hit.} to all subsystems including the TCP/IP stack.  For example, if we compare 4.15 performance to 5.3, we can see the latter having around $15\%$ higher goodput:
\begin{figure}[h]
	\includegraphics[width=3.31in]{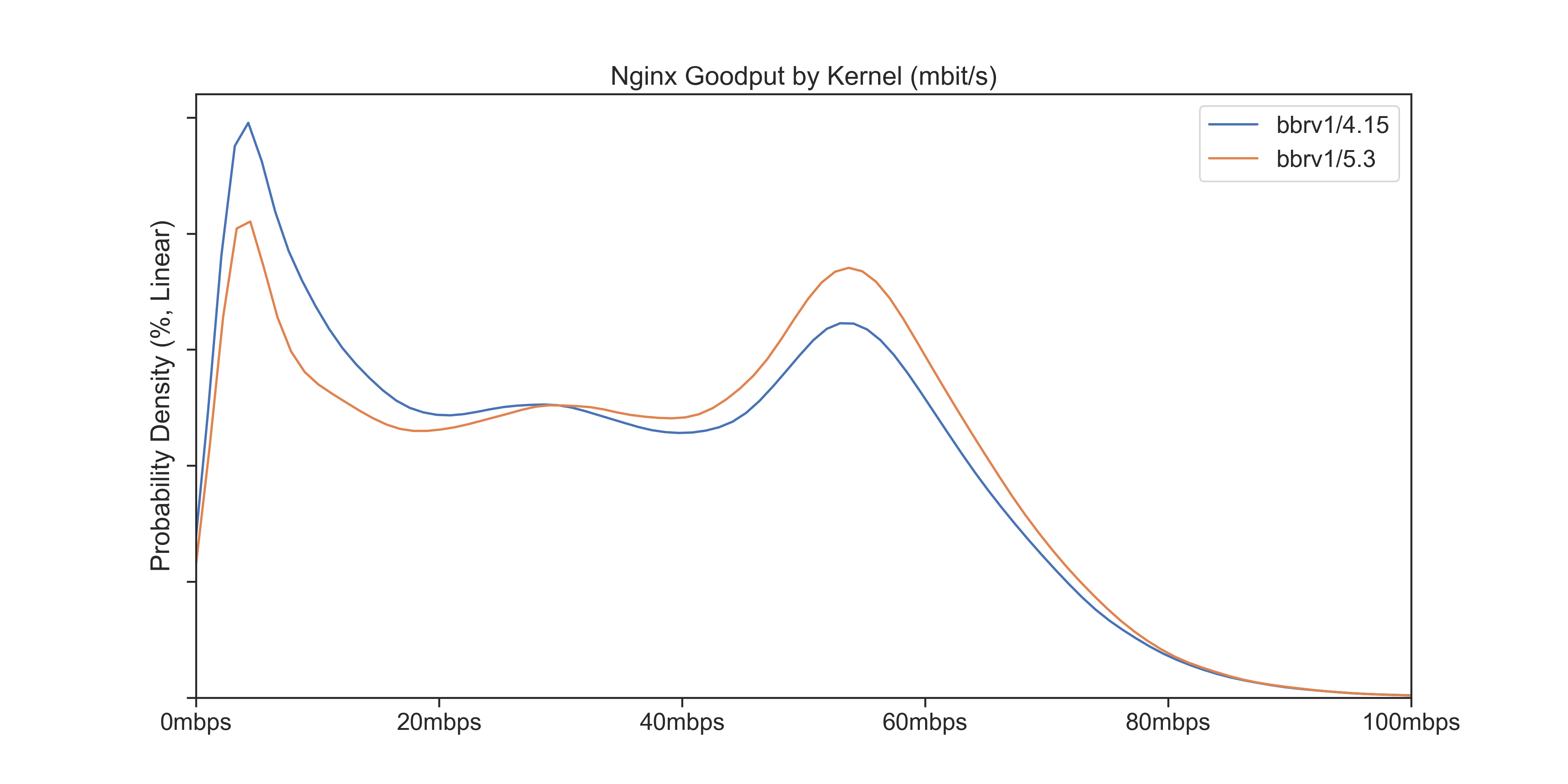}
	\caption{4.15 vs 5.3 kernels server-side file upload goodput.}
	\label{fig:kernelgoodput}
\end{figure}

Most likely candidates for this improvement are:

\begin{itemize}
\item ``tcp\_bbr: adapt cwnd based on ack aggregation estimation,'' fixing Wi-Fi performance.
\item ``tcp: switch to Early Departure Time model,'' fixing the RTT jitter observed by TCP when used with pacing (See ``\nameref{sec:afap}'' section.)
\end{itemize}

\subsection{Keeping userspace up-to-date}

Having recent versions of userspace is quite important if you are using kernels that are newer compared to what your OS was bundled with.  This is especially true for packages like \texttt{ethtool} and \texttt{iproute2}.

In our configuration we've used Ubuntu 16.04 Xenial with a fairly recent 5.3.0 kernel along with \texttt{iproute2-5.4.0}.  Here is an example of using the new version of \texttt{ss} (with the new output fields in \textbf{bold}):
\\
\begin{lstlisting}[
frame=single,
breaklines=true,
breakindent=0pt,
morekeywords={pmtu,rcvmss,minrtt,rwnd_limited,advmss,bytes_retrans,rcv_ssthresh,bw,mrtt,pacing_gain,cwnd_gain,delivery_rate},
caption={iproute2 5.4.0},
label=list:new-iproute,
]
$ ss -tie
ts sack bbr rto:220 rtt:16.139/10.041 ato:40 mss:1448 pmtu:1500 rcvmss:1269 advmss:1428 cwnd:106 ssthresh:52 bytes_sent:9070462 bytes_retrans:3375 bytes_acked:9067087 bytes_received:5775 segs_out:6327 segs_in:551 data_segs_out:6315 data_segs_in:12 bbr:(bw:99.5Mbps,mrtt:1.912,pacing_gain:1,cwnd_gain:2) send 76.1Mbps lastsnd:9896 lastrcv:10944 lastack:9864 pacing_rate 98.5Mbps delivery_rate 27.9Mbps delivered:6316 busy:3020ms rwnd_limited:2072ms(68.6%) retrans:0/5 dsack_dups:5 rcv_rtt:16.125 rcv_space:14400 rcv_ssthresh:65535 minrtt:1.907
\end{lstlisting}

As you can see, the new \texttt{ss} version has all the new goodies from the kernel’s \texttt{struct tcp\_info}, plus the internal BBR state from the \texttt{struct tcp\_bbr\_info}.  This adds lots of metrics that can be used even in day-to-day TCP performance troubleshooting.  Including very useful insufficient sender buffer and insufficient receiver window/buffer stats from the ``tcp: sender chronographs instrumentation'' patchset.

\section{Experimental results}

\subsection{Packet loss}

Most notably, switching from BBRv1 to BBRv2 results in an enormous drop in retransmits:

\begin{figure}[h]
	\includegraphics[width=3.31in]{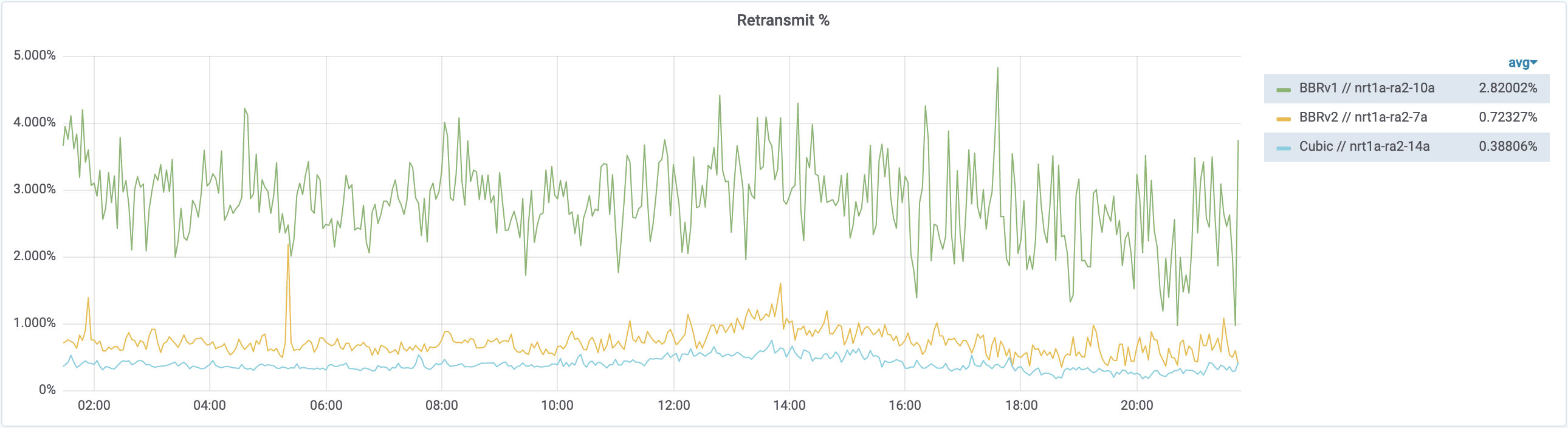}
	\caption{Retransmits, \%. CUBIC, BBRv1, and BBRv2.}
	\label{fig:bbrv2packetloss}
\end{figure}
 
\texttt{RetransSegs} on these boxes is still higher than on ones using CUBIC but given that BBR was designed to ignore non-congestion induced packet loss, we would assume that things work as intended.

\begin{figure}[ht]
	\includegraphics[width=3.31in]{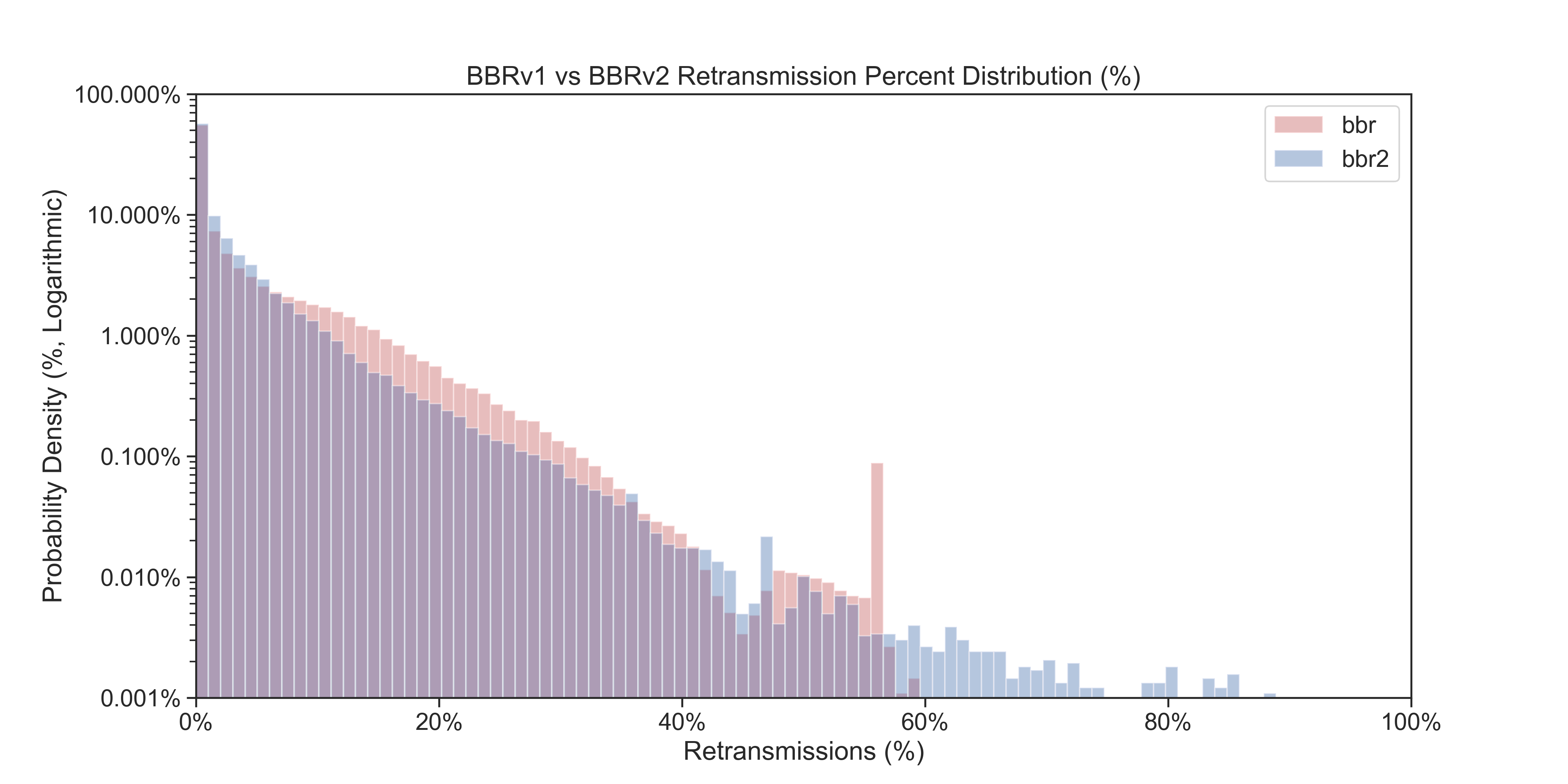}
	\caption{Retransmits, \%, PDF.  BBRv1 vs BBRv2.}
	\label{fig:retrans_bbr}
\end{figure}

\begin{figure}[ht]
	\includegraphics[width=3.31in]{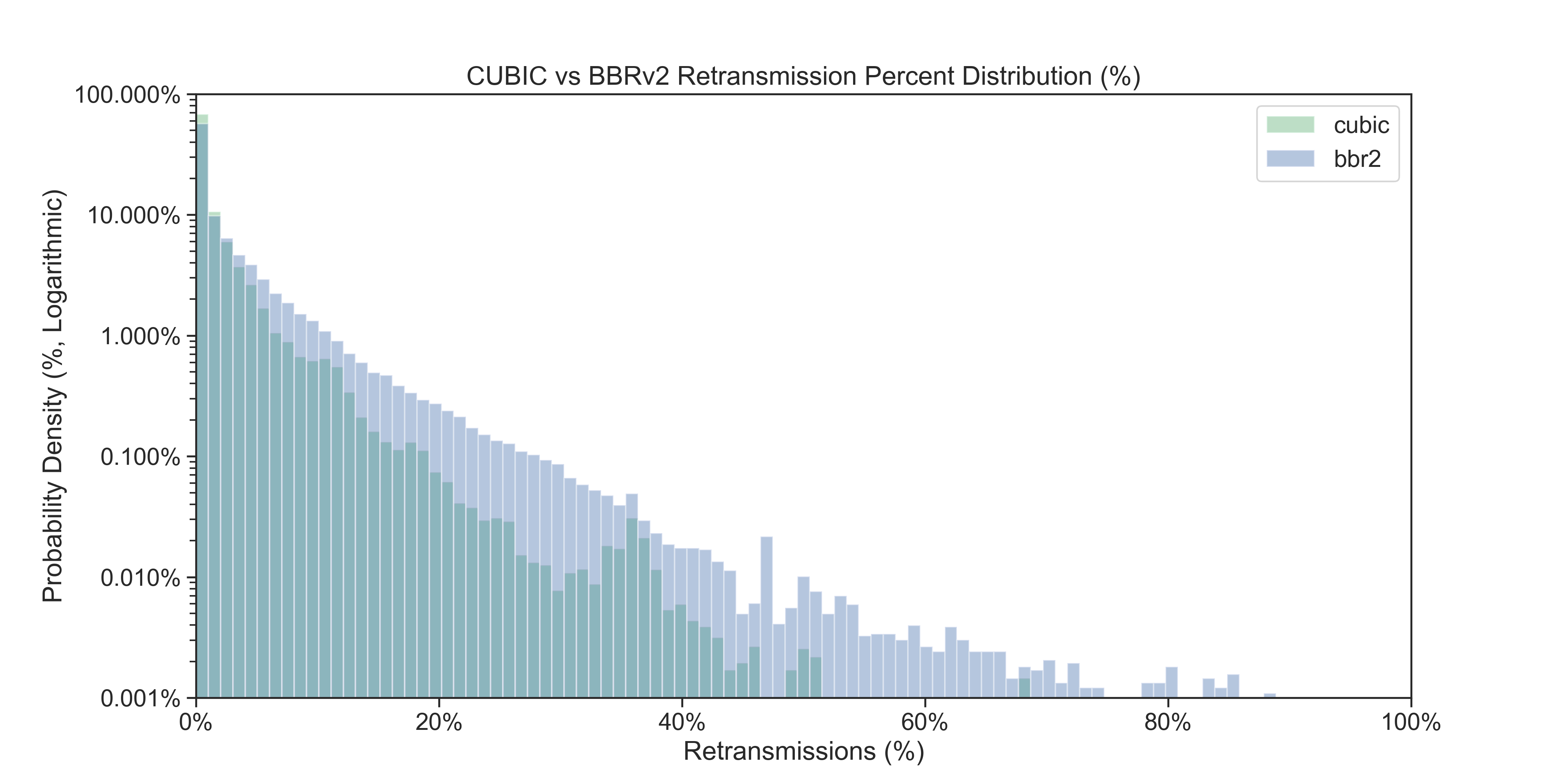}
	\caption{Retransmits, \%. PDF.  CUBIC vs BBRv2.}
	\label{fig:retrans_cubic}
\end{figure}

Looking deeper at the \texttt{ss} stats we can confirm this: BBRv2 packet loss is multiple times lower than BBRv1 (Figure~\ref{fig:retrans_bbr}.) though still higher than {CUBIC} (Figure~\ref{fig:retrans_cubic}.).

The deeper inspection also shows that BBRv2 has connections with $>60\%$ packet loss.  These types of outliers are present neither on BBRv1 nor CUBIC machines.  Looking at some of these connections closer does not reveal any obvious patterns: connections with absurdly large packet loss come from different OS’es (based on Timestamp/ECN support,) connections types (based on MSS,) and locations (based on RTT.)

Aside from these outliers, BBRv2 retransmissions are lower across all RTTs, when compared to BBRv1.

\begin{figure}[ht]
	\includegraphics[width=3.31in]{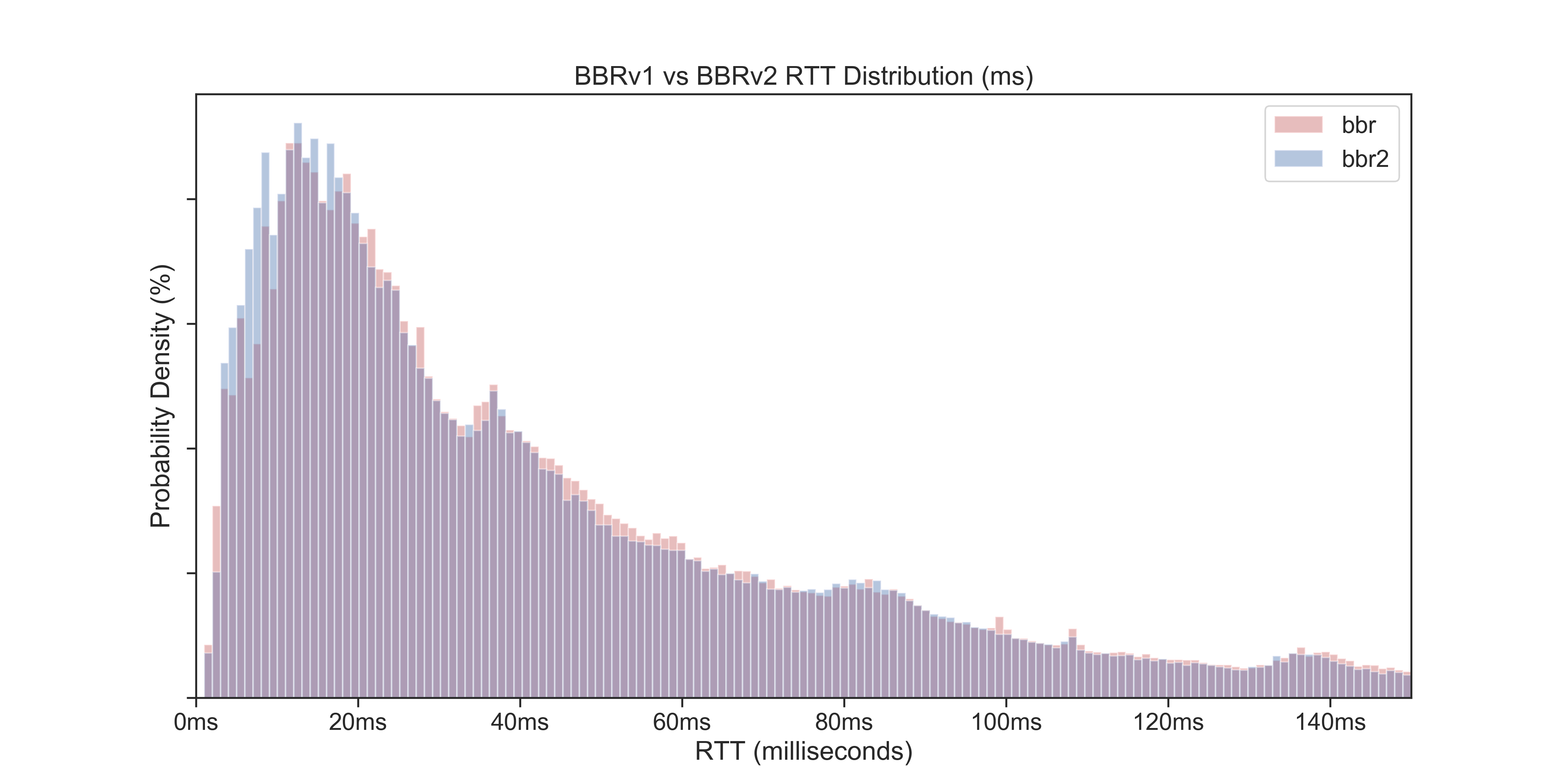}
	\caption{Round Trip Time.  BBRv1 vs BBRv2.}
	\label{fig:rtt_bbr}
\end{figure}

\subsection{Throughput}
\begin{figure}[ht]
	\includegraphics[width=3.31in]{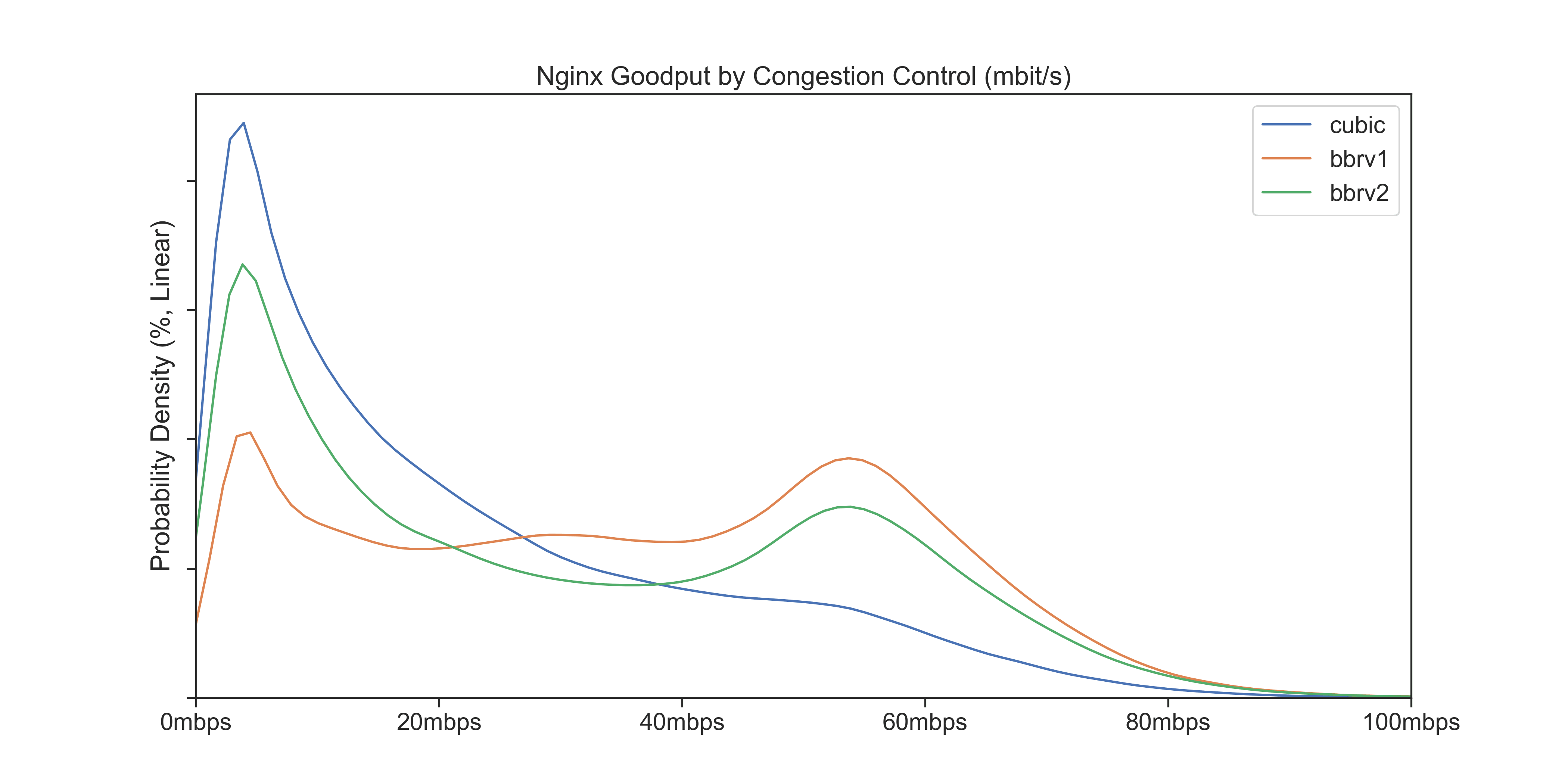}
	\caption{File upload goodput from nginx point of view.}
	\label{fig:nginxgoodput}
\end{figure}

On the throughput side, we looked at the nginx file upload goodput metric\footnote{From the Traffic Team's perspective, one of our SLIs for edge performance is end-to-end client-reported download speed.  For this test, we’ve been using server-side file upload speed as the closest proxy for it.} (Figure~\ref{fig:nginxgoodput}.). For lower percentiles of connection speeds, BBRv2 performance is closer to {CUBIC}. For higher ones, it starts getting closer to BBRv1.  This is likely due to BBRv2 being fairer to other TCP connections on the bottleneck since the slower a connection is the more likely it is being congested (instead of being app limited.)

\begin{figure}[ht]
	\includegraphics[width=3.31in]{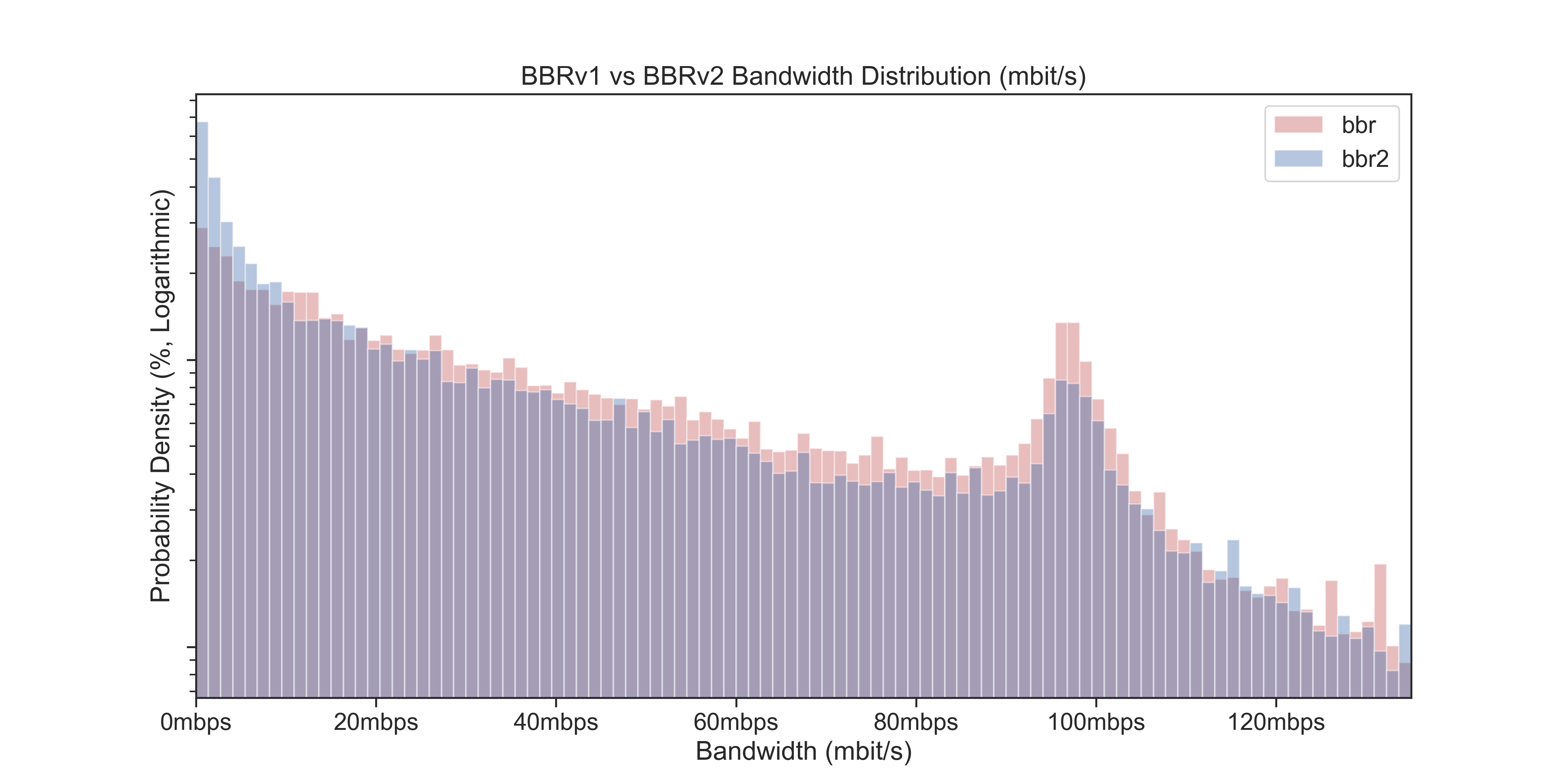}
	\caption{Bandwidth distribution.  BBRv1 vs BBRv2.}
	\label{fig:bw_bbr}
\end{figure}

\begin{figure}[ht]
	\includegraphics[width=3.31in]{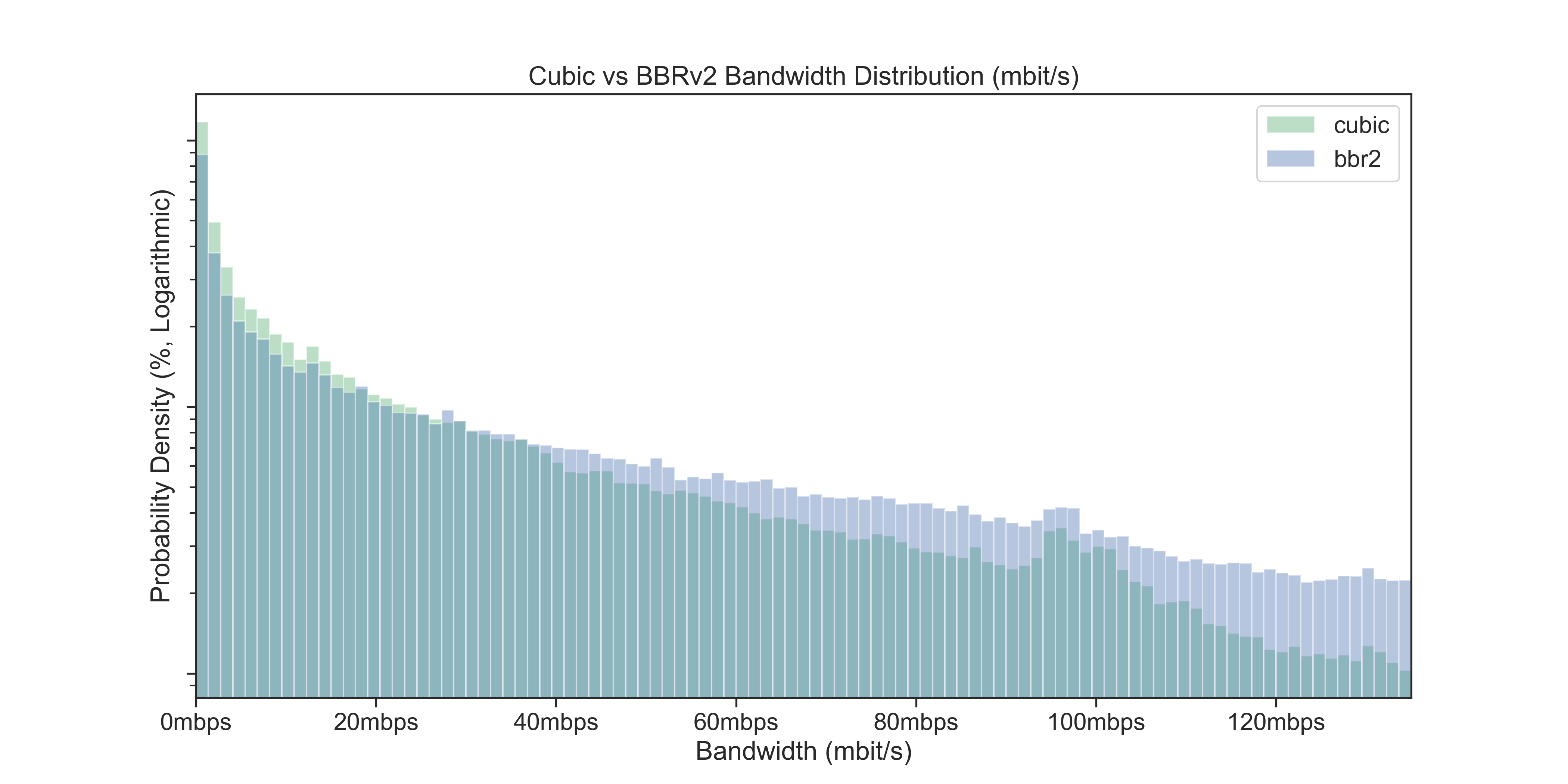}
	\caption{Bandwidth distribution.  CUBIC vs BBRv2}
	\label{fig:bw_cubic}
\end{figure}

Connection-level stats confirm that BBRv2 has lower bandwidth than BBRv1 (Figure~\ref{fig:bw_bbr}.) but still higher than CUBIC (Figure~\ref{fig:bw_cubic}.).

So, does that mean that BBRv2 is slower? Yes, it does, at least to some extent. So, what are we getting in return? Based on connection stats, quite a lot. We’ve already mentioned lower packet loss (and therefore higher ``goodput'',) but there is more.

\subsection{Packets in-flight}
\begin{figure}[ht]
	\includegraphics[width=3.31in]{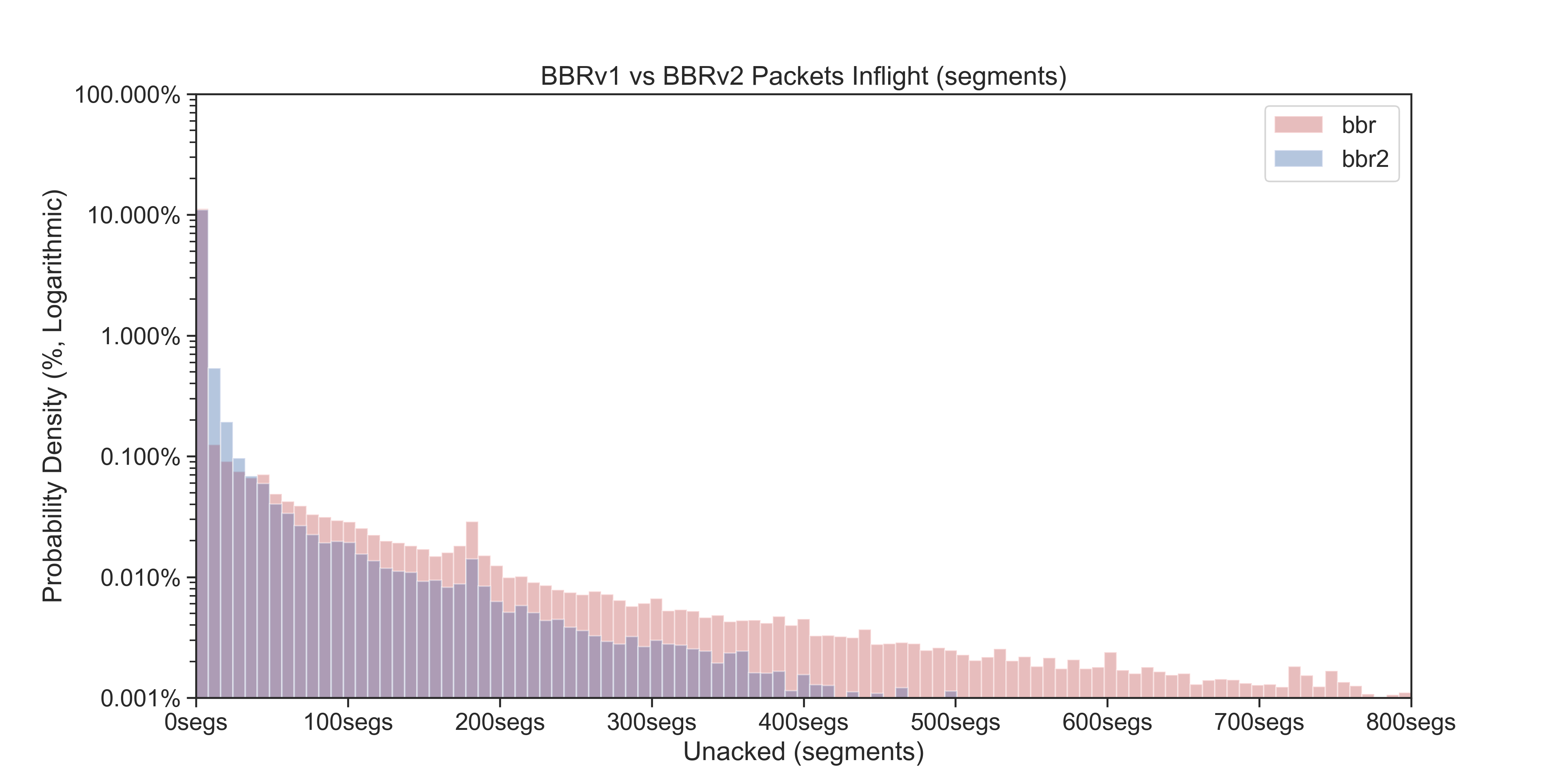}
	\caption{Unacked packets distribution.  BBRv1 vs BBRv2.}
	\label{fig:inflight_bbr}
\end{figure}

\begin{figure}[ht]
	\includegraphics[width=3.31in]{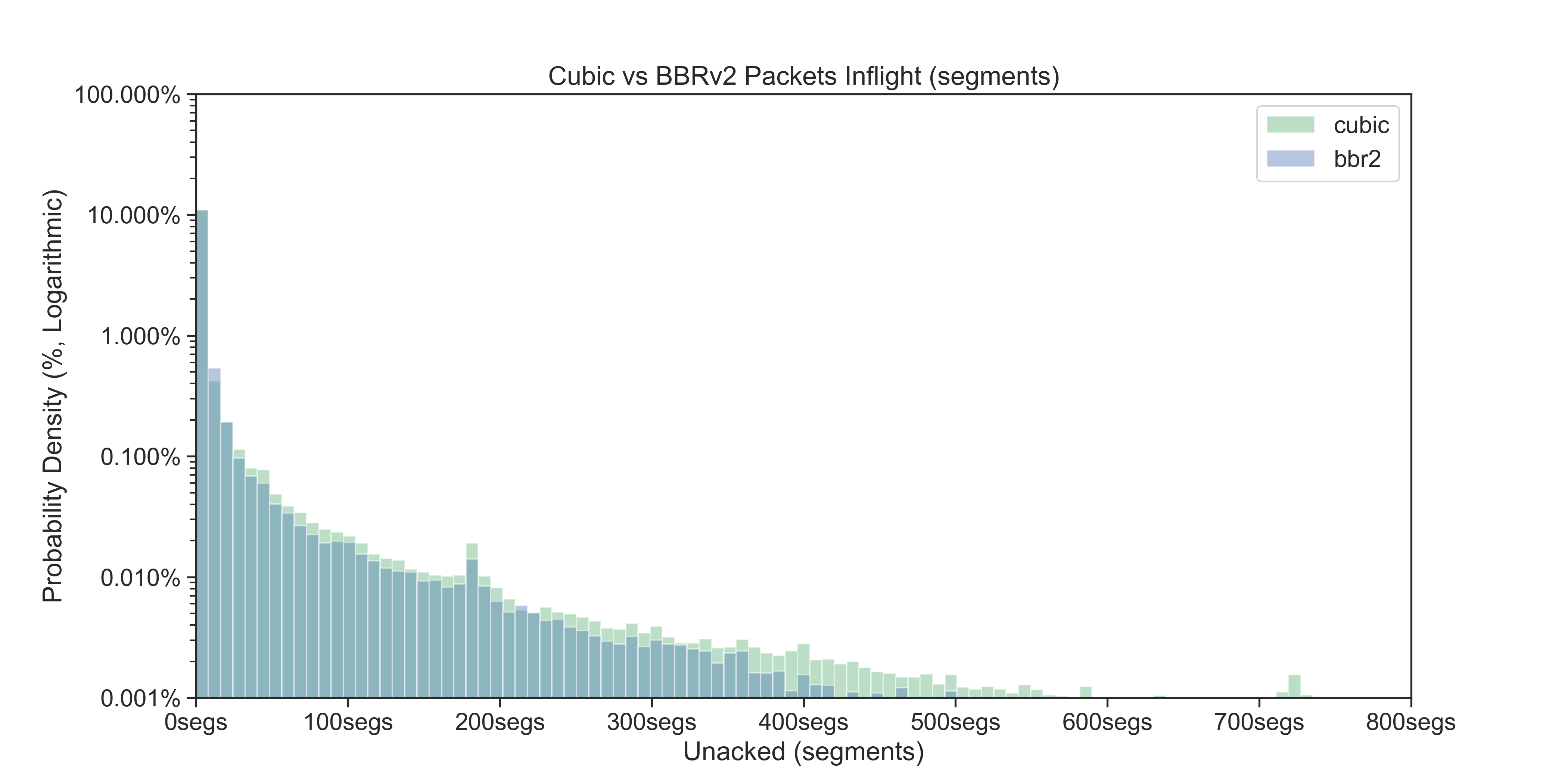}
	\caption{Unacked packets distribution.  CUBIC vs BBRv2.}
	\label{fig:inflight_cubic}
\end{figure}

We’ve observed way fewer “unacked” packets which is a good proxy for bytes in-flight. BBRv2 looks way better than in BBRv1 (Figure~\ref{fig:inflight_bbr}.) and even slightly better than CUBIC (Figure~\ref{fig:inflight_cubic}.).


\begin{figure}[ht]
	\includegraphics[width=3.31in]{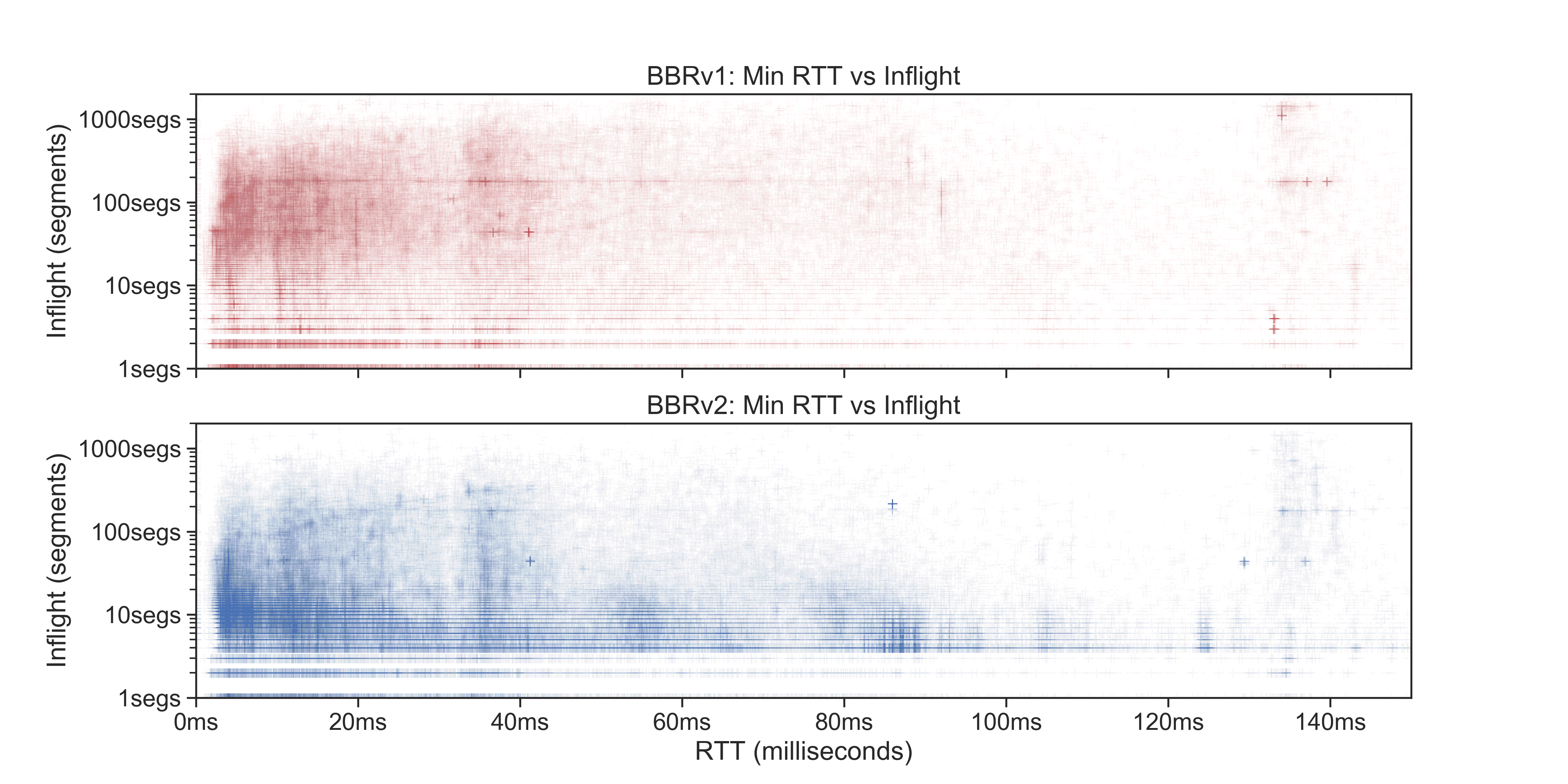}
	\caption{Unacked vs RTT heatmap.  BBRv1 vs BBRv2.}
	\label{fig:inflight_rtt}
\end{figure}

Plotting RTT vs in-flight heatmap shows that in the BBRv1 case amount of data in-flight tends to be dependent on the RTT.  This is fixed in BBRv2 (Figure~\ref{fig:inflight_rtt}.).

\begin{figure}[ht]
	\includegraphics[width=3.31in]{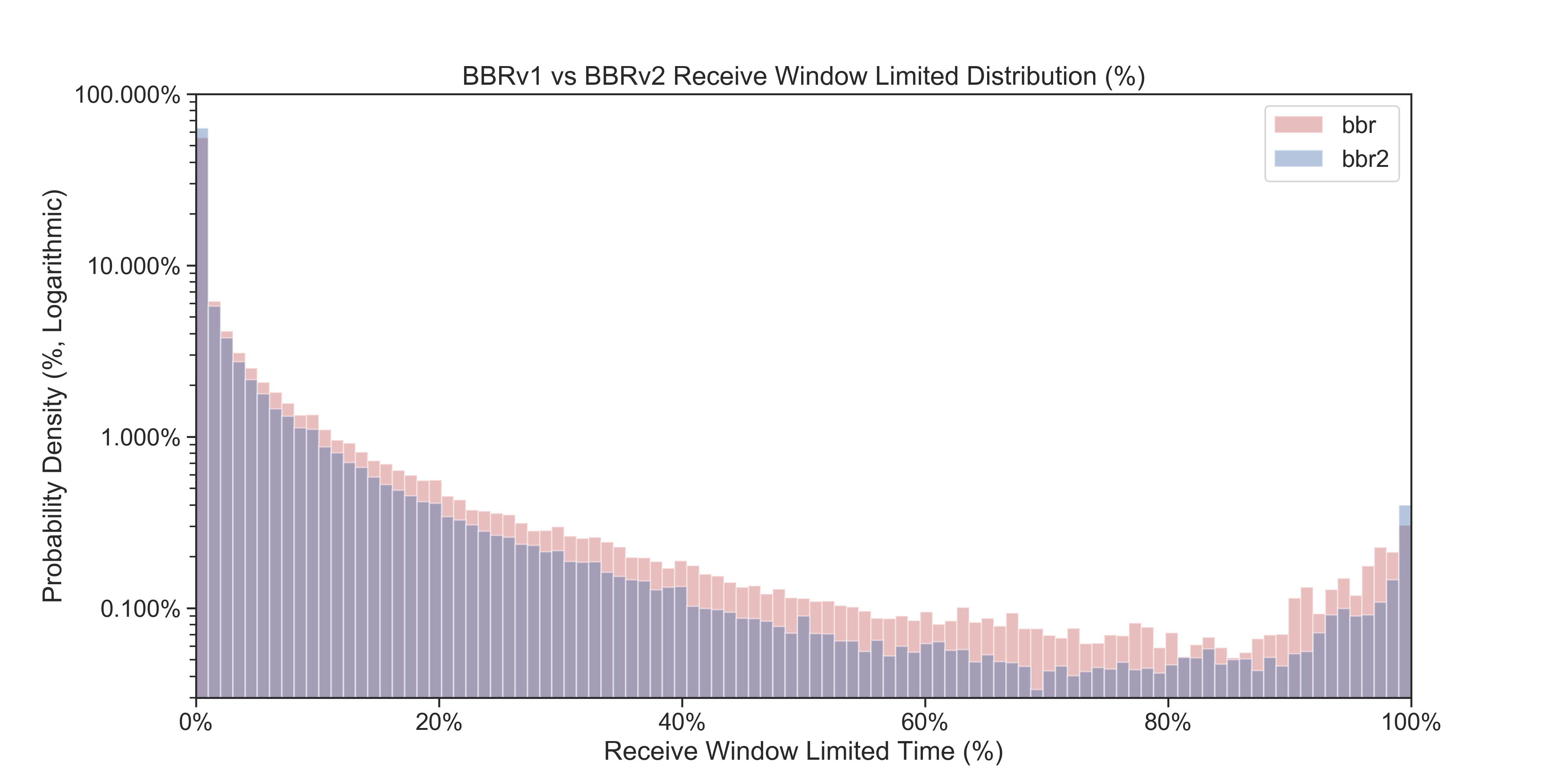}
	\caption{Receive Window Limited.  BBRv1 vs BBRv2.}
	\label{fig:rwnd_bbr}
\end{figure}

As BBR co-author, Neal Cardwell, explains it:
\begin{quote}
    ``In all of the cases I’ve seen with unfairness due to differing $min\_rtt$ values, the dominant factor is simply that with BBRv1 each flow has a cwnd that is basically $2*bw*min\_rtt$, which tends to try to maintain $1*bw*min\_rtt$ in the bottleneck queue, which quite directly means that flows with higher $min\_rtt$ values maintain more packets in the bottleneck queue and therefore get a higher fraction of the bottleneck bandwidth. The most direct way I’m aware of to improve RTT fairness in the BBR framework is to get rid of that excess queue, or ensure that the amount of queue is independent of a flow’s $min\_rtt$ estimate.''
\end{quote}

\subsection{Receive Window Limited connections}
\begin{figure}[ht]
	\includegraphics[width=3.31in]{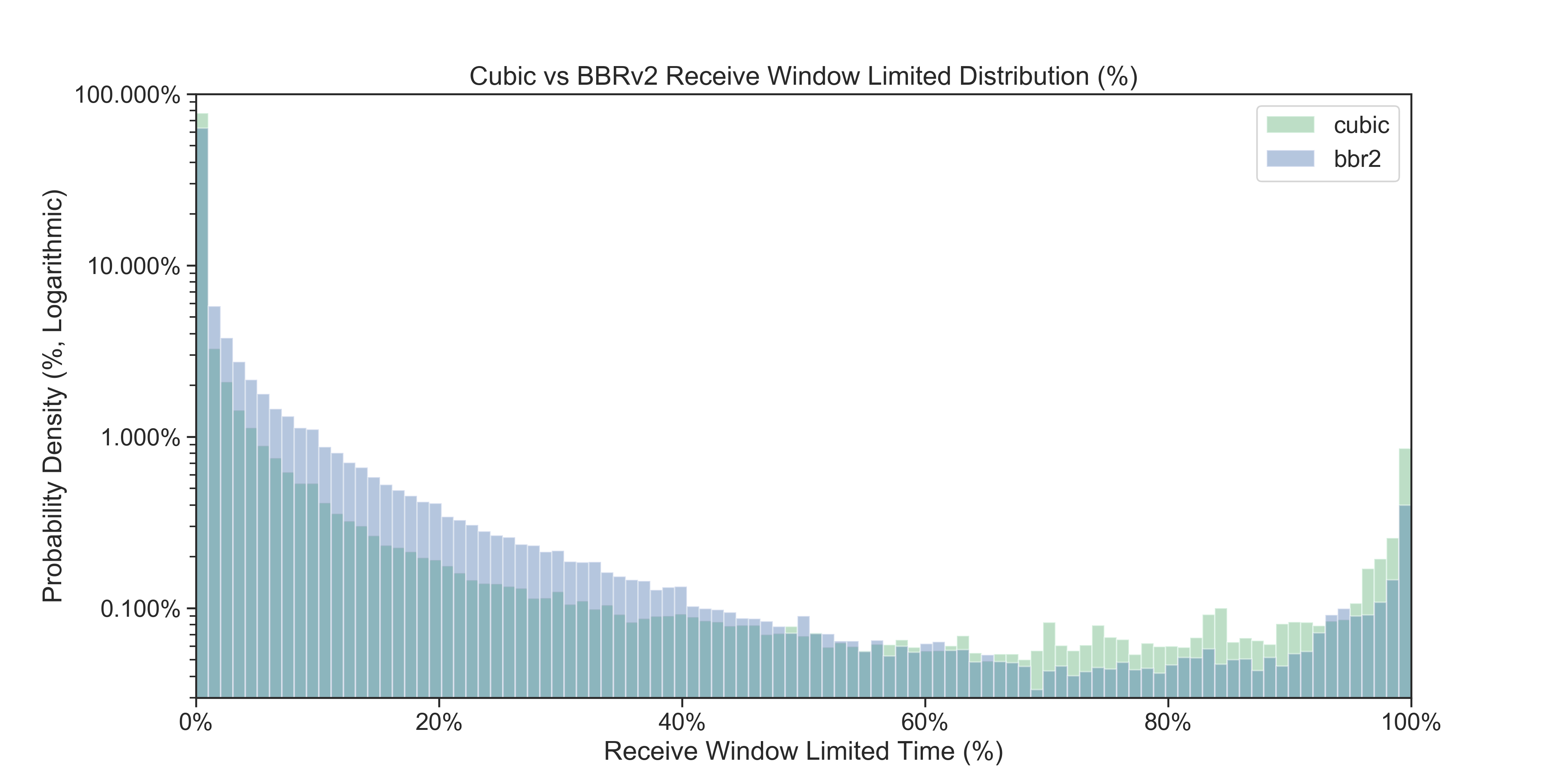}
	\caption{Receive Window Limited.  CUBIC vs BBRv2.}
	\label{fig:rwnd_cubic}
\end{figure}

We similarly observed that BBRv2 connections spend way less time being receive window limited than both BBRv1 (Figure~\ref{fig:rwnd_bbr}.) and CUBIC\footnote{CUBIC was using \texttt{sch\_fq} and pacing in this test too.}(Figure~\ref{fig:rwnd_cubic}.).

\begin{figure}[ht]
	\includegraphics[width=3.31in]{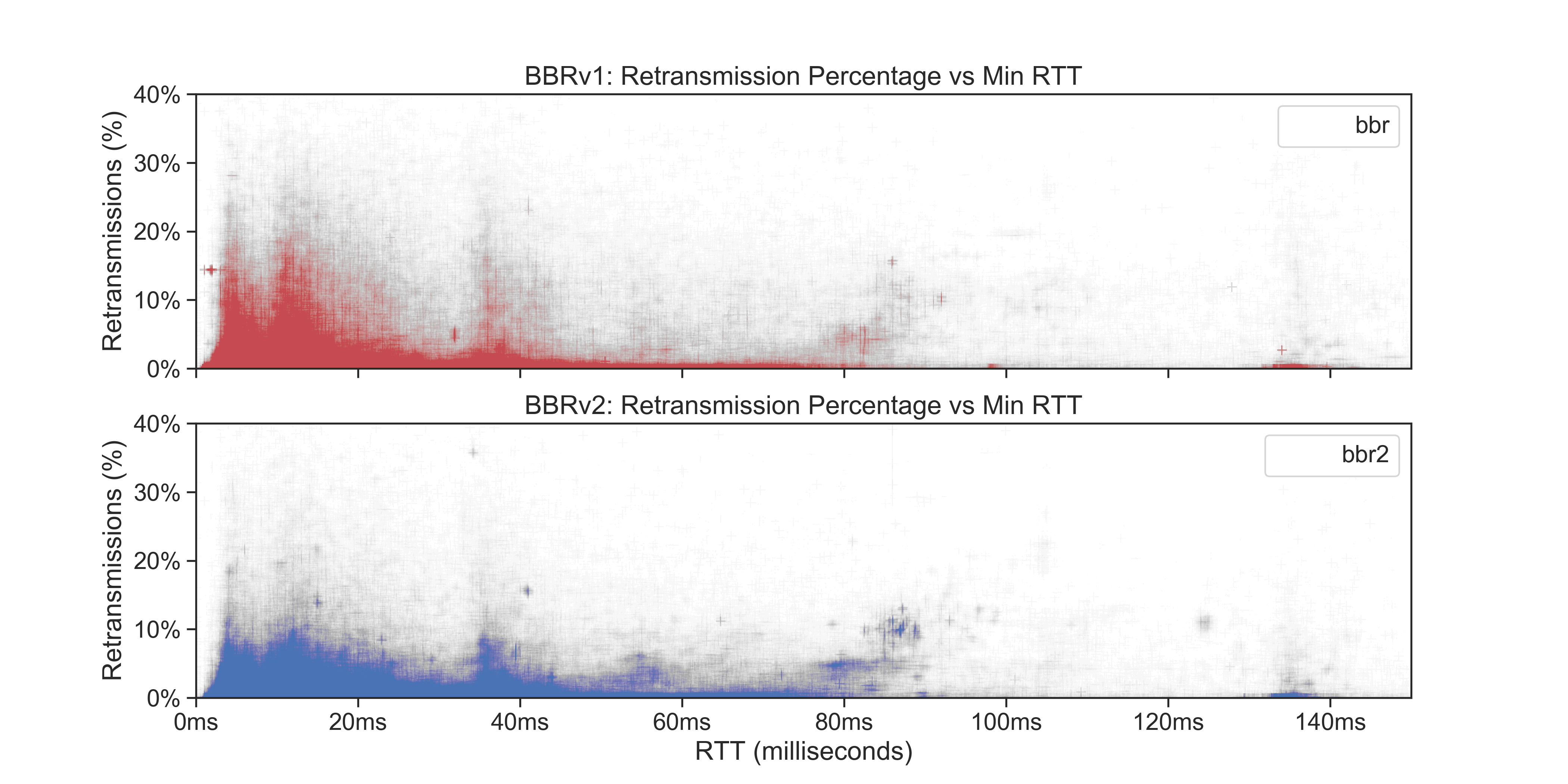}
	\caption{Retransmits vs RTT heatmap.  BBRv1 vs BBRv2.}
	\label{fig:retrans_vs_rtt}
\end{figure}

\subsection{Round Trip Time}
Based on connection stats BBRv2 also has a lower RTT than BBRv1 (Figure~\ref{fig:rtt_bbr}.).  This can be explained by a more graceful \texttt{PROBE\_RTT} phase.

\subsection{Correlations between Min RTT and Bandwidth}
\begin{figure}[ht]
	\includegraphics[width=3.31in]{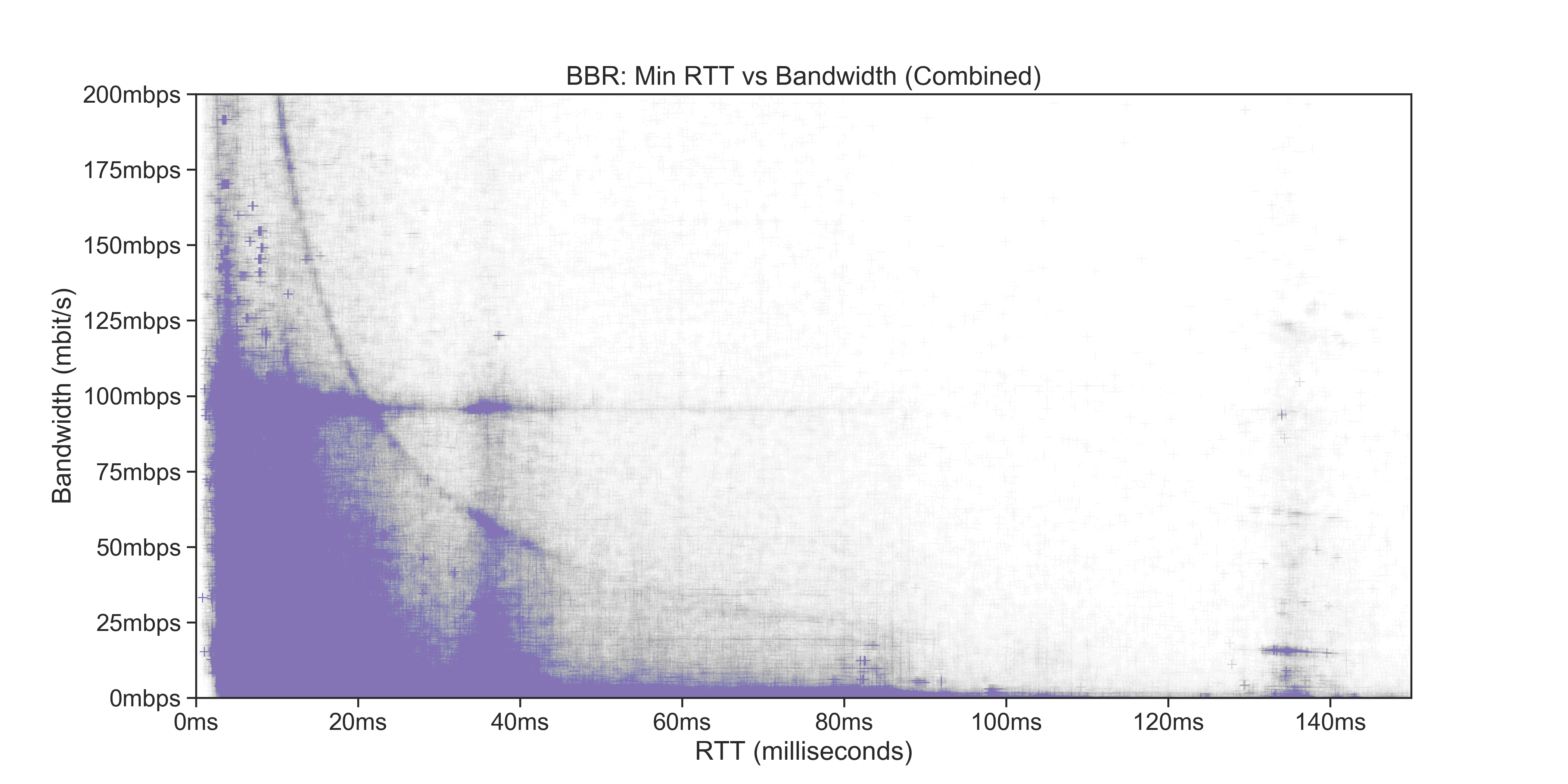}
	\caption{\texttt{bbr\_mrtt} vs \texttt{bbr\_bw} heatmap.}
	\label{fig:mrtt_bw}
\end{figure}

If we construct \texttt{bbr\_mrtt} vs \texttt{bbr\_bw} heatmap then vertical bands represent a network distance from user to our Tokyo PoP\footnote{The 130ms RTT band represents cross-Pacific traffic and can be attributed to GSLB failure to properly route users to the closest PoP. This since have been fixed by RUM DNS\cite{Shirokov2020}\cite{Guba2018}.} while horizontal bands represent common {Internet} speeds (Figure~\ref{fig:mrtt_bw}.)

What’s interesting here is the exponentially decaying relationship between MinRTT and bandwidth for both BBRv1 and BBRv2.  This means that there are some cases where bandwidth is artificially limited by RTT.  Since this behavior is the same between BBRv1 and BBRv2 we did not dig too much into it.  It’s likely though that the bandwidth is being artificially limited by the user’s receive window.

\subsection{CPU Usage}

Previously, BBRv1 updated the whole model on each ACK received, which is quite a lot of work given that our average server receives millions of them every second. BBRv2 has an even more sophisticated network model but it also adds a fast path for ACK processing\footnote{Along with ACK fast-path other optimizations to the BBR were introduced, aimed at both CPU usage improvements and goodput increase e.g. improved TSO auto-sizing, faster ACKs, and reduced standing queuing at the bottleneck with many competing BBR flows\cite{ietf106}.} that skips model updates for the application-limited case. This, in theory, should greatly reduce CPU usage on common workloads.

In our tests, though, we did not see any measurable difference in CPU usage between BBRv1 and BBRv2. This is likely due to BBRv2 having quite a lot of debug code enabled.

\section{Conclusions}

In our testing, BBRv2 showed the following properties:
\begin{itemize}
\item Bandwidth is comparable to CUBIC for users with lower percentiles of Internet speeds.
\item Bandwidth is comparable to BBRv1 for users with higher percentiles of Internet speeds.
\item Packet loss is 4 times lower compared to BBRv1\footnote{Minus the 0.0001\% of the outliers with a $>60\%$ packet loss.}; still 2 times higher than CUBIC.
\item Data in-flight is 3 times lower compared to BBRv1; slightly lower than CUBIC.
\item RTTs are lower compared to BBRv1; still higher than CUBIC.
\item Higher RTT-fairness compared to BBRv1.
\end{itemize}
Overall, BBRv2 is a great improvement over the BBRv1 and indeed seems way closer to being a \textbf{drop-in replacement for Reno/CUBIC} in cases where one needs higher bandwidth with lower buffer bloat. Adding the experimental ECN support to that and we can even see BBRv2 being used as a \textbf{drop-in replacement for Data Center TCP (DCTCP)}.

\appendix
\appendixpage

\section{Explicit Congestion Notification}
\label{sec:ecn}

ECN is a mechanism for the network bottleneck to proactively notify the sender to slow down before it runs out of buffers and starts “tail dropping” packets.  Currently, though, ECN on the Internet is mostly deployed in a so-called “passive” mode. Based on Apple’s data $74\%$ most popular websites “passively” support ECN\cite{Cheshire2017}.  In our Tokyo PoP, \textbf{we observe $3.68\%$ of connections being negotiated with ECN, $88\%$ of which have the \texttt{ecnseen} flag}.

One downside of the Classic ECN\cite{rfc3168} is that it's too prescriptive about the explicit congestion signal.  Some RFCs, like the “Problem Statement and Requirements for Increased Accuracy in Explicit Congestion Notification (ECN) Feedback\cite{rfc7560},” call out the low granularity of classic ECN that is only able to feed back one congestion signal per RTT.  Also for a good reason: DCTCP (and BBRv2\footnote{Pay special attention to the CPU usage if you are testing BBR with ECN enabled since it may render GRO/GSO/TSO unusable for high packet loss scenarios.}) implementation of ECN greatly benefits from its increased accuracy\cite{Alizadeh2010}.

Another RFC, namely “Relaxing Restrictions on Explicit Congestion Notification (ECN) Experimentation\cite{rfc8311},” tries to fix it by relaxing this requirement.  That way implementations are free to choose behavior outside of the one specified by Classic ECN.

Talking about ECN it’s hard to not mention that there is also a “congestion-notification conflict” going over a single code point (a half a bit) of space in the IP header between the “Low Latency, Low Loss, Scalable Throughput (L4S)”\cite{ietf-tsvwg-l4s-arch-04} proposal and the bufferbloat folks behind the “The Some Congestion Experienced ECN Codepoint (SCE)”\cite{morton-tsvwg-sce-01} draft.

As Jonathan Corbet summarizes it:
\begin{quotation}
    ``These two proposals are clearly incompatible with each other; each places its own interpretation on the \texttt{ECT(1)} value and would be confused by the other. The SCE side argues that its use of that value is fully compatible with existing deployments, while the L4S proposal turns it over to private use by suitably anointed protocols that are not compatible with existing congestion-control algorithms. L4S proponents argue that the dual-queue architecture is necessary to achieve their latency objectives; SCE seems more focused on fixing the endpoints.''
\end{quotation} 

Time will show which, if any, draft is approved by IETF, in the meantime, we can all help the Internet by deploying AQMs\cite{rfc7567}(e.g. fq-codel\cite{rfc8290}, cake\cite{HoilandJorgensen2018}) to the network bottlenecks under our control.

\section{Beyond As Fast As Possible}
\label{sec:afap}

\subsection{Evolving from AFAP – Teaching NICs about time}
There is a great talk by Van Jacobson about the evolution of computer networks and that sending “as fast as possible” is not an optimal strategy in today’s Internet and even inside a datacenter\cite{Jacobson2018}.

This talk is a great summary of the reasons why one might consider using pacing on the network layer and a delay-based congestion control algorithm.

\subsection{Fair Queue scheduler}
All our Edge boxes are running Fair Queue qdisc.  Our main goal is not the fair queuing itself but the pacing introduced by \texttt{sch\_fq}\footnote{Earlier fq implementations did add some jitter to TCP’s RTT estimations which can be problematic inside the data center since it will likely inflate p99s of RPC requests.  This was solved in ``tcp: switch to Early Departure Time model.'' and should be available since Linux 4.20.}.

One can use \texttt{bpftrace qdisc-fq.bt}\footnote{\texttt{qdisc-fq.bt} is a part of supplementary material to the ``BPF Performance Tools: Linux System and Application Observability'' book by Brendan Gregg\cite{Gregg2019}.} to measure the time difference between packets being enqueued into the qdisc and dequeued from it, and hence the effect of pacing on the network.

\begin{lstlisting}[
frame=single,
caption={\texttt{qdisc-fq.bt} output on a live system.},
label=list:bpftrace,
]
# bpftrace qdisc-fq.bt
@us:
[0]               237486 |               |
[1]              8712205 |@@@@           |
[2, 4)          21855350 |@@@@@@@@@@@@   |
[4, 8)           4378933 |@@             |
[8, 16)           372762 |               |
[16, 32)          178145 |               |
[32, 64)          279016 |               |
[64, 128)         603899 |               |
[128, 256)       1115705 |               |
[256, 512)       2303138 |@              |
[512, 1K)        2702993 |@@             |
[1K, 2K)        11999127 |@@@@@@@        |
[2K, 4K)         5432353 |@@@            |
[4K, 8K)         1823173 |@              |
[8K, 16K)         778955 |               |
[16K, 32K)        385202 |               |
[32K, 64K)        146435 |               |
[64K, 128K)        31369 |               |
[128K, 256K)        2967 |               |
[256K, 512K)         271 |               |
\end{lstlisting}

In our tests, deploying FQ across the Dropbox internal network essentially eliminated the premium queue frame discards on shallow-buffered switches.

\onecolumn{
    \bibliographystyle{isea}
    \bibliography{isea}
}

\end{document}